\documentclass[prb,11pt,onecolumn]{revtex4}

\usepackage{graphicx,amssymb,amsmath,pdfpages}

\begin{document}

\title{Surface Recombination Limited Lifetimes of Photoexcited Carriers in Few-Layer Transition Metal Dichalcogenide MoS$_{2}$}
\author{Haining Wang}
\email{hw343@cornell.edu}
\author{Changjian Zhang}
\author{Farhan Rana}
\affiliation{School of Electrical and Computer Engineering, Cornell University, Ithaca, NY, USA}

\begin{abstract}

We present results on photoexcited carrier lifetimes in few-layer transition metal dichalcogenide MoS$_2$ using nondegenerate ultrafast optical pump-probe technique. Our results show a sharp increase of the carrier lifetimes with the number of layers in the sample. Carrier lifetimes increase from few tens of picoseconds in monolayer samples to more than a nanosecond in 10-layer samples. The inverse carrier lifetime was found to scale according to the probability of the carriers being present at the surface layers, as given by the carrier wavefunction in few layer samples, which can be treated as quantum wells. The carrier lifetimes were found to be largely independent of the temperature and the inverse carrier lifetimes scaled linearly with the photoexcited carrier density. These observations are consistent with defect-assisted carrier recombination, in which the capture of electrons and holes by defects occurs via Auger scatterings. Our results suggest that carrier lifetimes in few-layer samples are surface recombination limited due to the much larger defect densities at the surface layers compared to the inner layers. 
\end{abstract}


\maketitle

\begin{figure}[tbh]
  \begin{center}
   \includegraphics[width=1.0\textwidth]{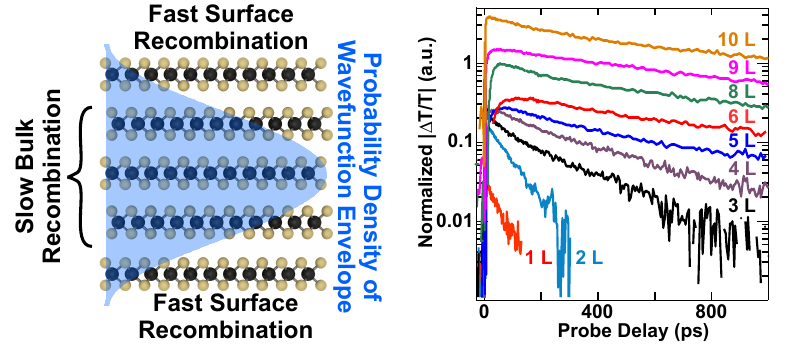}
    \label{pump}
  \end{center}
\end{figure}

\hfill \\
\noindent
\textbf{Introduction:}
\newline
Two dimensional (2D) transition metal dichalcogenides (TMDs) are proving to be interesting materials for a variety of different low-cost optoelectronic device applications, including photodetectors, light-emitting diodes, and, more recently, lasers~\cite{Mak10,Splendiani10,Wang12,Mak13,Lopez13,Ross14,Hua12,Steiner13,Baugher14,Changjian14,Sanfeng15}. Indirect bandgap few layer and bulk TMDs are well suited for high quantum efficiency photodetection applications because of their large optical absorption coefficients and long photoexcited carrier lifetimes~\cite{Strait14,Huang13,Yu13}. Few layer TMDs and their heterojunctions are promising for flexible photovoltaic and solar cell devices with thickness-adjustable bandgaps~\cite{Ma14,Hui14,Cho14,Yu13}. For all the aforementioned applications, understanding the dynamics of photoexcited carriers in these materials is important.  

The ultrafast dynamics of photoexcited carriers and the non-radiative recombination mechanisms in monolayer TMDs, and in MoS$_{2}$ in particular, have been the subject of several recent experimental~\cite{Wang15,Huang13,WangR12,Choi13,Sun14,Lagarde14,Schuller11,Docherty14} and theoretical studies~\cite{Wang15b}. Non-radiative electron-hole recombination in MoS$_{2}$ monolayers results in photoexcited carrier lifetimes in the few tens of picosecond range. In contrast, photoexcited carrier lifetimes in bulk MoS$_{2}$ have been shown to be in the few nanosecond range~\cite{Strait14,Huang13}. This large difference between the carrier lifetimes in monolayer and multilayer 2D materials remains poorly understood. Shi et~al. attributed this difference to the changes in the electronic bandstructure of MoS$_{2}$ going from monolayer, to few-layer, and to bulk~\cite{Huang13}.

Monolayer MoS$_{2}$ is known to have several different kinds of point defects, such as sulfur and molybdenum vacancies, interstitials, and adsorped impurity atoms, in addition to grain boundaries and dislocations~\cite{Sofo04,Komsa12,Seifert13,Zhou13,Kim14,Guinea14,Robertson13,Hao13,VanDerZande13}. Recently, defect-assisted electron-hole recombination was proposed as the dominant non-radiative recombination mechanism in monolayer and bulk MoS$_{2}$~\cite{Wang15,Strait14,Wang15b, Marco14}. The carrier density and the temperature dependence of the observed recombination dynamics suggested that photoexcited carriers are captured by defects via Auger processes~\cite{Wang15,Strait14,Wang15b}. It is reasonable to expect that surface layers of few-layer TMDs have far more defects and impurities than the inner layers and, consequently, if defect-assisted processes are indeed responsible for electron-hole recombination in TMDs, then the recombination time would scale in some meaningful way with the number of layers in few-layer TMDs.    

In this letter, we report results for ultrafast carrier dynamics from non-degenerate optical pump-probe studies of few-layer MoS$_{2}$. Our results show that the photoexcited carrier lifetimes increase dramatically from $\sim$50 ps in monolayer MoS$_2$ to $\sim$1 ns in 10-layer MoS$_{2}$. The lifetimes were found to be largely temperature independent in all few-layer MoS$_2$ samples irrespective of the number of layers. The evolution of the carrier lifetime with the number of layers matches extremely well with our analytical model. The analytical model assumes a fast recombination time for the two surface layers in a few-layer sample, and a slow recombination time for all the inner layers, and then estimates the actual recombination time for a few-layer sample by weighing the inverse lifetime with the probability of electron occupation of each layer as given by the electron wavefunction in a few-layer sample. The good agreement between the measurements and the model shows that electron-hole recombination in few-layer MoS$_{2}$ is dominated by defect-assisted recombination processes in which the surface layers play an important role. In addition, the temperature and pump fluence dependence of the measured lifetimes are consistent with carrier capture by defects via Auger scatterings~\cite{Wang15b,Wang15,Strait14}. Reduction of carrier lifetimes in more traditional semiconductor nanostructures, such as quantum wells and wires, due to very large surface recombination velocities is well known~\cite{Tsang88,Germann89,Moison90,Shalish04,Klimov99}, and surface passivation schemes have proven to be critical in the operation of optoelectronic devices based on these materials~\cite{Bright86,Wang13,Dan11}. Our work underscores the importance of developing similar passivation schemes for optoelectronic devices based on monolayer and multilayer TMDs.

\begin{figure}[tbh]
  \centering
  \includegraphics[width=1.0\textwidth]{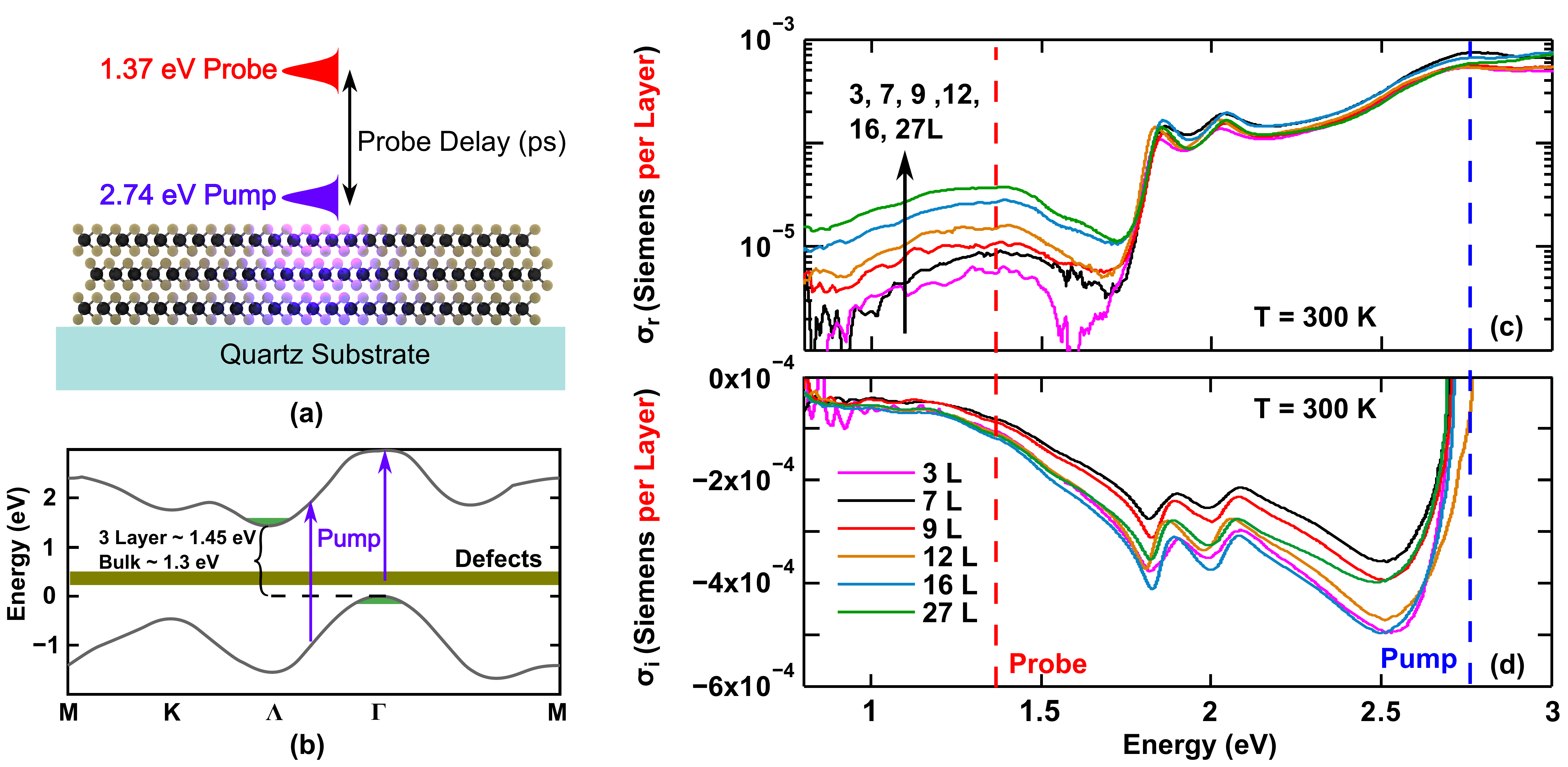}
  \caption
      {(a) Nondegenerate ultrafast optical-pump optical-probe (OPOP) experiment on few-layer MoS$_2$. (b) Bandstructure of few-layer MoS$_2$ and depiction of electron photoexcitations from the valence band and midgap defect states into the conduction bands. (c)(d) The measured optical conductivities (real part(c) and imaginary part(d)) of few-layer MoS$_{2}$ samples are normalized to their layer numbers and plotted.}
  \label{fig:OPOP1}
\end{figure}

\hfill \\
\noindent
\textbf{Sample Preparation and Experimental Technique:}
\newline
Few-layer MoS$_2$ samples were mechanically exfoliated from bulk MoS$_2$ crystals (obtained from 2D Semiconductors Inc.) and transferred onto quartz substrates. Exfoliated samples were characterized electrically and optically, using transmission/reflection spectroscopies, to determine the electronic and optical conductivities~\cite{Wang15}. The monolayer samples were found to be moderately n-doped ($>10^{12}$ 1/cm$^{2}$) and the multilayer samples were found to be lightly n-doped (2-4$\times 10^{15}$ 1/cm$^{3}$), consistent with previously reported results~\cite{Ryu10,Strait14,Wang15,Changjian14,Sood12}. The thickness of the samples was measured by AFM to determine the number of layers~\cite{Mak10}. In the nondegenerate ultrafast optical pump-probe (OPOP) experiments, $\sim$80 fs pulses at 905 nm center wavelength (1.37 eV photon energy) from a 83 MHz repetition rate Ti-Sapphire laser were frequency-doubled to 452 nm (2.74 eV) by a beta-BaB$_{2}$O$_{4}$ crystal. The 452 nm pump pulses were used to excite electrons from the valence band into the conduction band in the samples. The differential transmission ($\Delta T/T$) of time-delayed 905 nm probe pulses was measured using a chopped lock-in technique. A 20X objective was used to focus the pump and the probe pulses onto the samples. The OPOP experiment is illustrated in Figure \ref{fig:OPOP1}(a). Bandstructure of multilayer MoS$_2$ is shown in Figure \ref{fig:OPOP1}(b) which depicts the carrier excitation process~\cite{Mak10,Zhao13}. The measured optical conductivities (real part) for different layer numbers, are normalized to the layer number and shown in Figure \ref{fig:OPOP1}(c). Real and imaginary part of the optical conductivity are extracted from combined transmission and reflection measurements and the details are discussed in the supplementary information. The interband optical conductivity (above $\sim$1.9 eV) is seen to scale almost linearly with the number of layers. A weak absorption band was observed in all multilayer samples in the 0.8-1.6 eV range. Interestingly, this absorption band was not observed in monolayer samples~\cite{Wang15}. We rule out indirect interband absorption as solely responsible for this absorption band because the band extends to energies much smaller than the indirect bandgap even in our thickest samples~\cite{Ellis11,Mak10}. Since all our samples were n-doped, this absorption could be due to optical transitions from midgap defect states to the conduction bands or due to optical transitions from the lowest conduction band to higher conduction bands. From the optical conductivity measurements, 1 $\mu$J/cm$^{2}$ pump fluence is estimated to generate an electron (and hole) density that varies from $\sim$2.5$\times$10$^{11}$ 1/cm$^{2}$ in monolayer to $\sim$1.5$\times$10$^{11}$ 1/cm$^{2}$ per layer in 10-layer MoS$_{2}$. The measurement time resolution was $\sim$300-350 fs, and was limited by the dispersion of the optics in the setup. The maximum probe delay was limited by our setup to $\sim$1 ns. The goal of the experiment was to use the free-carrier absorption of the probe pulse by the photoexcited electrons and holes to monitor their temporal dynamics. The same technique was used by the authors to study the ultrafast dynamics of photoexcited carriers in monolayer MoS$_{2}$~\cite{Wang15}.       

\begin{figure}[tbh]
  \centering
  \includegraphics[width=1.0\textwidth]{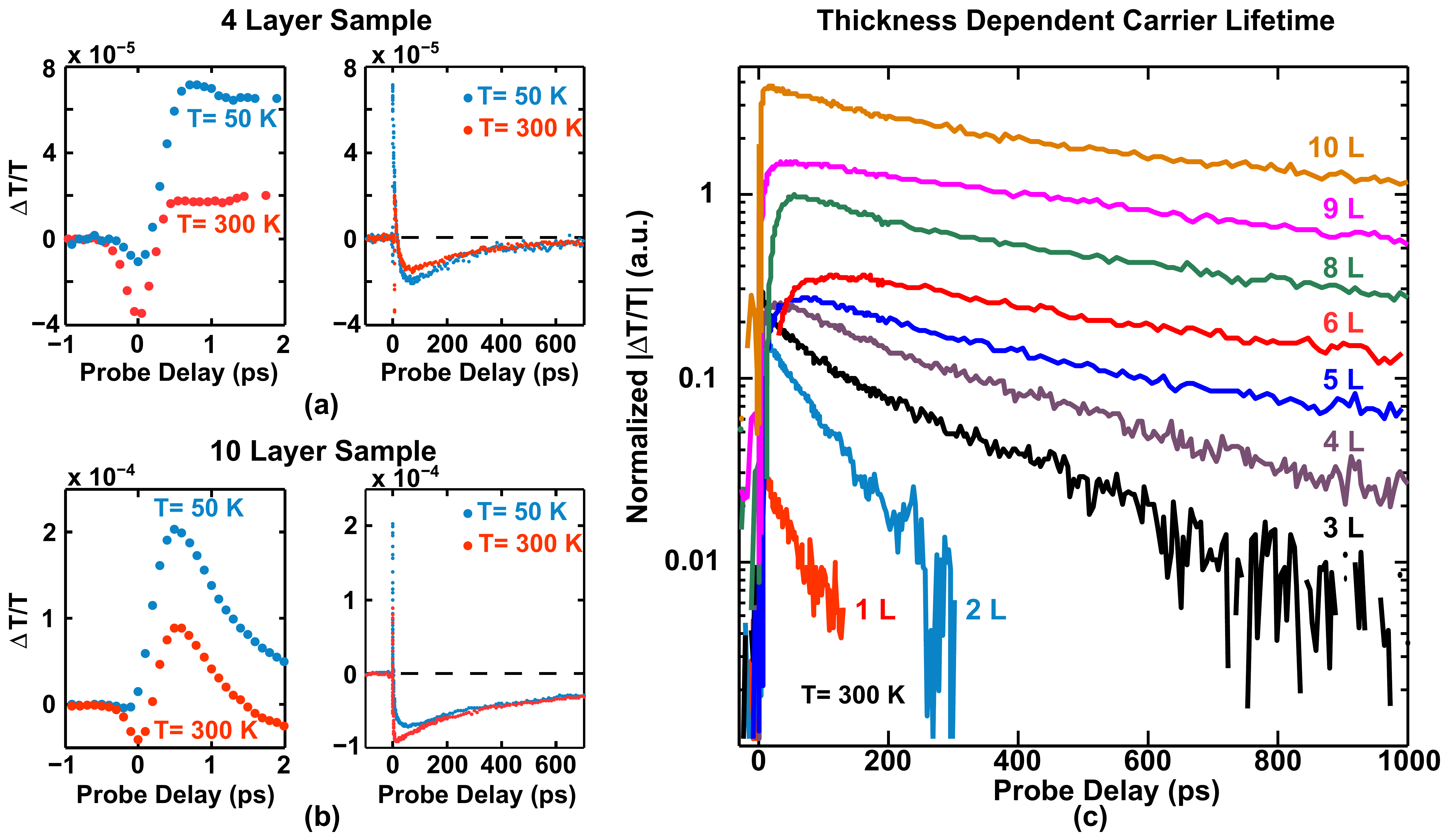}
  \caption
      {(a,b) Measured differential transmission transients ($\Delta T/T$) of few-layer MoS$_2$ ((a) 4-layer and (b) 10-layer as representatives of all measured samples) are plotted at different time scales and for different substrate temperatures. The negative dip in $\Delta T/T$ near zero probe delay is attributed to two-photon absorption between the pump and probe pulses. The positive part of $\Delta T/T$ right after photoexcitation comes from the decreased defect absorption of the probe pulse caused by the ionization of the defects by the pump pulse. The long negative part of $\Delta T/T$, which lasts from tens of picoseconds to nanoseconds, is due to the intraband absorption of the probe pulse by the photoexcited carriers. The time scales exhibited in the long negative part of $\Delta T/T$ are temperature independent. (c) The magnitude of long negative $\Delta T/T$ transients are plotted on a log scale to show the dependence of the time scales on the layer number. Different curves are scaled in magnitude for clarity. Pump fluence was 32 $\mu$J/cm$^{2}$ in all measurements.}
  \label{fig:OPOP2}
\end{figure}

\hfill \\
\noindent
\textbf{Experimental Results and Discussion:}
\newline
Measured differential transmission transients $\Delta T/T$ for few-layer samples (4-layer and 10-layer) are shown in Figure \ref{fig:OPOP2}(a)-(b) for two different temperatures. The results for monolayer samples were reported in previous work~\cite{Wang15}, and the results shown in Figure \ref{fig:OPOP2}(a)-(b) are representative of all multilayer samples studied in this work. The results for multilayer samples exhibit the following three prominent features: (a) Near zero probe delay, $\Delta T/T$ dips negative for a duration shorter than $\sim$0.5 ps. We attribute this dip to two-photon absorption between the pump and probe pulses when they overlap. The duration and shape of the dip are both consistent with this interpretation. The magnitude of the dip is very sensitive to the degree of overlap between the pump and probe pulses. (b) $\Delta T/T$ then immediately turns positive for a duration that can last anywhere from a few picoseconds to more than ten picoseconds and the duration was found to have no meaningful dependence on the layer number. This portion of the transient is temperature dependent, becoming larger at smaller temperatures, and is absent in monolayer samples~\cite{Wang15}. (c) Finally, $\Delta T/T$ turns negative and this last portion of the transient exhibits time scales that are independent of the temperature and vary from tens of picoseconds to more than a nanosecond, becoming longer for samples with larger number of layers. These time scales can be seen more clearly in Figure \ref{fig:OPOP2}(c) which plots the magnitude of the transients on a log scale (after rescaling them for clarity). The observed time scales exhibit an interesting relationship with the number of layers in the sample and this relationship is discussed in detail later in the paper. 

We first discuss different physical mechanisms contributing to $\Delta T/T$ in our measurements. The measured differential transmission $\Delta T/T$ of the probe pulse can be expressed as~\cite{Wang15},
\begin{equation} 
\frac{\Delta T}{T} \approx - \frac{\displaystyle 2 \eta_{o} \frac{\Delta\sigma_{r}}{1+n_{s}} + 2 \eta_{o}^{2} \frac{\sigma_{r}\Delta\sigma_{r} + \sigma_{i}\Delta\sigma_{i}}{(1+n_{s})^{2}}}{\displaystyle \left| 1 +  \eta_{o} \frac{\sigma_{r}+i\sigma_{i}}{1+n_{s}} \right|^{2}} \label{eq:DeltaT}
\end{equation}  
where $\sigma_{r}$($\sigma_{i}$) is the real(imaginary) part of intraband optical conductivity (units: Siemens), $n_{s}$ is the refractive index of SiO$_{2}$ and $\eta_{o}$ is the vacuum impedance. The positive part of the transient is attributed to the decreased probe absorption by the optically active midgap defect states due to pump induced ionization of these defect states, and it is described by a negative value of $\Delta\sigma_{r}$. The rather long relaxation times associated with the positive part of the transient rule out bleaching of the probe intraband absorptions by the pump pulse as the cause of the positive part of the transient. Another mechanism that can contribute to a negative value of $\Delta\sigma_{r}$ is Pauli blocking of the indirect interband absorption of the probe pulse by the photoexcited carriers. The latter is ruled out because a strong negative contribution to the value of $\Delta\sigma_{r}$ is observed even in bilayer and trilayer samples in which the measured indirect bandgaps (1.6 eV for bilayer and 1.45 eV for trilayer~\cite{Mak10}) are much larger than the 1.37 eV probe photon energy. In addition, the positive part of the $\Delta T/T$ transient is absent in monolayer samples~\cite{Wang15}, which is consistent with no measurable midgap defect absorption in monolayer MoS$_2$~\cite{Wang15}. The larger magnitudes of the positive part of $\Delta T/T$ at lower temperatures, seen in  Figure \ref{fig:OPOP2}(a)-(b), are attributed to the larger initial occupation of the midgap defect states by electrons in our lightly n-doped materials at lower temperatures and this is consistent with a larger measured value of the real part of the optical conductivity for these midgap defects at lower temperatures (see supplementary information). The final negative part of the transient is attributed to intraband absorption of the probe pulse by the photoexcited carriers and is described by a positive value of $\Delta\sigma_{r}$. The Drude component of the intraband absorption by the photoexcited carriers is described by the conductivity change~\cite{Wang15},
\begin{equation}
  \Delta\sigma_{r} + i\Delta \sigma_{i} \approx e^{2} \left( \frac{1}{\omega^{2} \tau_{d}} + \frac{i}{\omega} \right) \,  \left( \frac{\Delta n}{m_{e}} + \frac{\Delta p}{m_{h}} \right) \label{eq:sigma}
\end{equation}
Here, $\Delta n$ ($\Delta p$) is the photoexcited electron (hole) density in the conduction (valence) band, $\omega$ is the probe optical frequency, and $\tau_{d}$ is the carrier momentum scattering time (assumed to be the same for both electrons and holes). The final negative part of the transient is related directly to the density of the photoexcited electrons and holes in the bands and enables one to measure the time scales associated with electron-hole recombination, as discussed in detail below.


The refractive index response is given by the changes in the imaginary part of the optical conductivity which can affect the probe transmission through the term containing the product $\sigma_{i}\Delta\sigma_{i}$ in Eq.(\ref{eq:DeltaT}). The measured imaginary part of the optical conductivity $\sigma_{i}$ of MoS$_{2}$ for different number of layers is shown in Figure \ref{fig:OPOP1}(d). $\sigma_{i}<0$ at the probe wavelength. Since the intraband (see Eq.(\ref{eq:sigma})) as well as the interband contributions to $\Delta \sigma_{i}$ from the photoexcited free-carriers are both positive~\cite{scsm67,tld98}, the contribution to $\Delta T/T$ from $\Delta \sigma_{i}$ will also be positive which is not consistent with the long negative part of the measured $\Delta T/T$ transient. Note that the magnitude of the product $\eta_{o} \sigma_{i}$, which appears with $\Delta \sigma_{i}$ in Eq.(\ref{eq:DeltaT}), is much less than unity at the probe wavelength for all MoS$_{2}$ samples considered in this work. Therefore, the contribution from the real part of the conductivity, as given by the first term on the right hand side in Eq.(\ref{eq:DeltaT}), dominates the measured response. Finally, one also needs to consider the effect of the temperature on the refractive index or, equivalently, on the imaginary part of the optical conductivity. In most semiconductors, an increase in the temperature results in an increase in the refractive index and a decrease (more negative) in the imaginary part of the conductivity ~\cite{Modine94,tld98}. Consequently, an increase in the sample temperature by the pump pulse would result in a negative contribution to the value of $\Delta T/T$. If the measured negative part of the $\Delta T/T$ transient in our experiments were due to temperature relaxation, then the associated time scale, determined by the ratio of the material heat capacity and the relevant thermal conductivity, would be independent of the pump fluence. However, as we show below, the measured time scales depend on the pump fluence consistent with defect-assisted recombination of photoexcited carriers via Auger scatterings~\cite{Wang15}. Therefore, temperature relaxation can also be ruled out as a dominant contributing factor to the measured $\Delta T/T$ transients. 

\begin{figure}[tbh]
  \centering
  \includegraphics[width=1.0\textwidth]{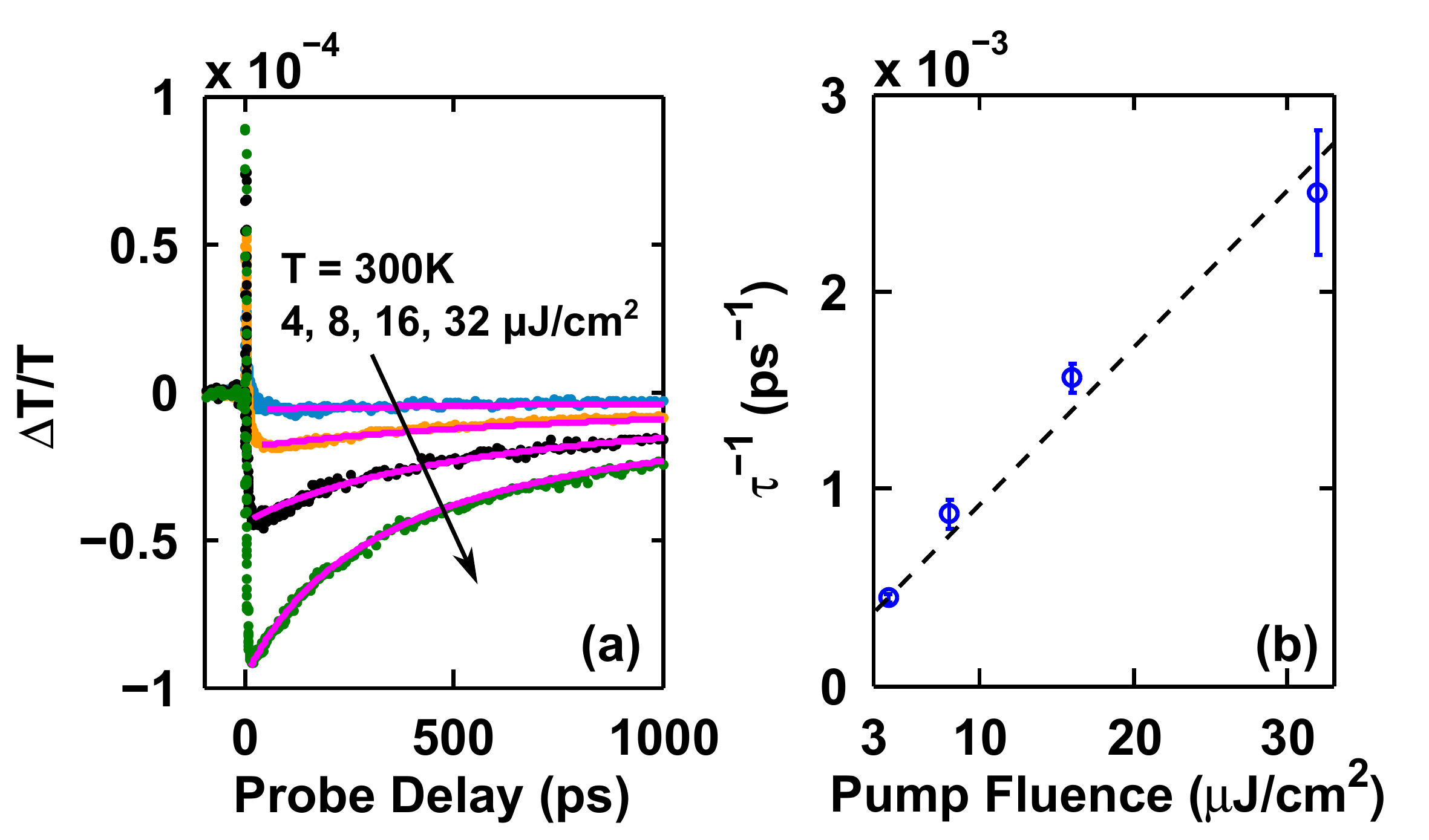}
  \caption
      {(a) (Dots) Measured differential transmission transients $\Delta T/T$ of a 10-layer MoS$_2$ sample are plotted for different values of the pump fluence (4, 8, 16, 32 $\mu$J/cm$^{2}$). T=300K. (Solid lines) Theoretical model for electron-hole recombination by defect-assisted processes in which carrier capture by defects occurs via Auger scattering. (b) The observed inverse carrier lifetime $1/\tau$ right after photoexcitation scales linearly with the photoexcited carrier density and, therefore, with the pump fluence. The recombination model fits the data very well. The vertical offset in the data at zero pump fluence is attributed to the light n-doping in our sample.}
  \label{fig:fitting}
\end{figure}

\hfill \\
\noindent
\textbf{Model for Electron-Hole Recombination, Carrier Lifetimes, and Comparison with Data:}
\newline
Models for electron-hole recombination (involving free-carriers as well as excitons) by defect-assisted processes in which carrier capture by defects occurs via Auger scattering have been presented by the authors in previous works~\cite{Wang15,Wang15b,Strait14}, and used successfully to model the carrier recombination dynamics in monolayer and bulk MoS$_{2}$ samples~\cite{Wang15,Strait14}. A prominent feature of Auger scattering is recombination times that are independent of the temperature but depend on the carrier density (or the pump fluence). In all the MoS$_{2}$ samples studied in this work, we observe this feature in the long negative $\Delta T/T$ transients. Therefore, following our earlier work~\cite{Wang15,Strait14}, we model the recombination dynamics with electron and hole defect capture rates, $R_{e}$ and $R_{h}$, respectively, that are given by the following expressions valid for our n-doped samples: $R_{e} = An^{2}n_{d}(1-F_{d})$ and $R_{h}  = B n p n_{d}F_{d}$. Here, $n$ ($p$) is the electron (hole) density, $n_{d}$ is the defect density, $F_{d}$ is the defect occupation probability, and $A$ and $B$ are rate constants for Auger scattering. Details of the model can be found in the supplementary information. Figure \ref{fig:fitting}(a) shows the good agreement between the recombination model and the measured $\Delta T/T$ transients for the long negative part of the transients for different pump fluence values. If the the photoexcited carrier density is much larger than the defect density $n_{d}$ and the equilibrium carrier density, then it can be shown that for multilayer samples the carrier recombination dynamics can be described by the following Equation (see the supplementary information),
\begin{equation}
  \frac{dn}{dt} \approx \frac{dp}{dt} \approx  - \frac{(An_{d})(Bn_{d})}{An_{d} + Bn_{d}} \, np
  \end{equation}
It is the product of the rate constant and the defect density that determines the recombination rates. According to the above Equation, the initial inverse carrier lifetime $1/\tau$ after photoexcitation is expected to scale linearly with the photoexcited carrier density and, therefore, with the pump fluence. The extracted initial inverse carrier lifetimes $1/\tau$ plotted in Figure \ref{fig:fitting}(b) are seen to scale almost linearly with the pump fluence, as expected from the model. The vertical offset in Figure \ref{fig:fitting}(b) at zero pump fluence is attributed to the light n-doping in our samples. Note that none of our measurements rule out or affirm if the defects responsible for electron-hole recombination are the same midgap defects whose signature appears in the optical absorption spectra discussed above. Photoexcited carrier dynamics in monolayer MoS$_{2}$ are not adequately described by the above Equation~\cite{Wang15}. In the case of monolayer MoS$_{2}$, fast defects contribute to rapid electron-hole recombination in the first few picoseconds and carrier recombination dynamics on long time scales are described by slow defects. A more detailed discussion of the carrier dynamics in monolayer samples is given previously by Wang et~al.~\cite{Wang15}. In Figure \ref{fig:OPOP2}(c), only the portion of the transient governed by the slow defects is visible for the monolayer sample.

\begin{figure}[tbh]
  \includegraphics[width=.4\textwidth]{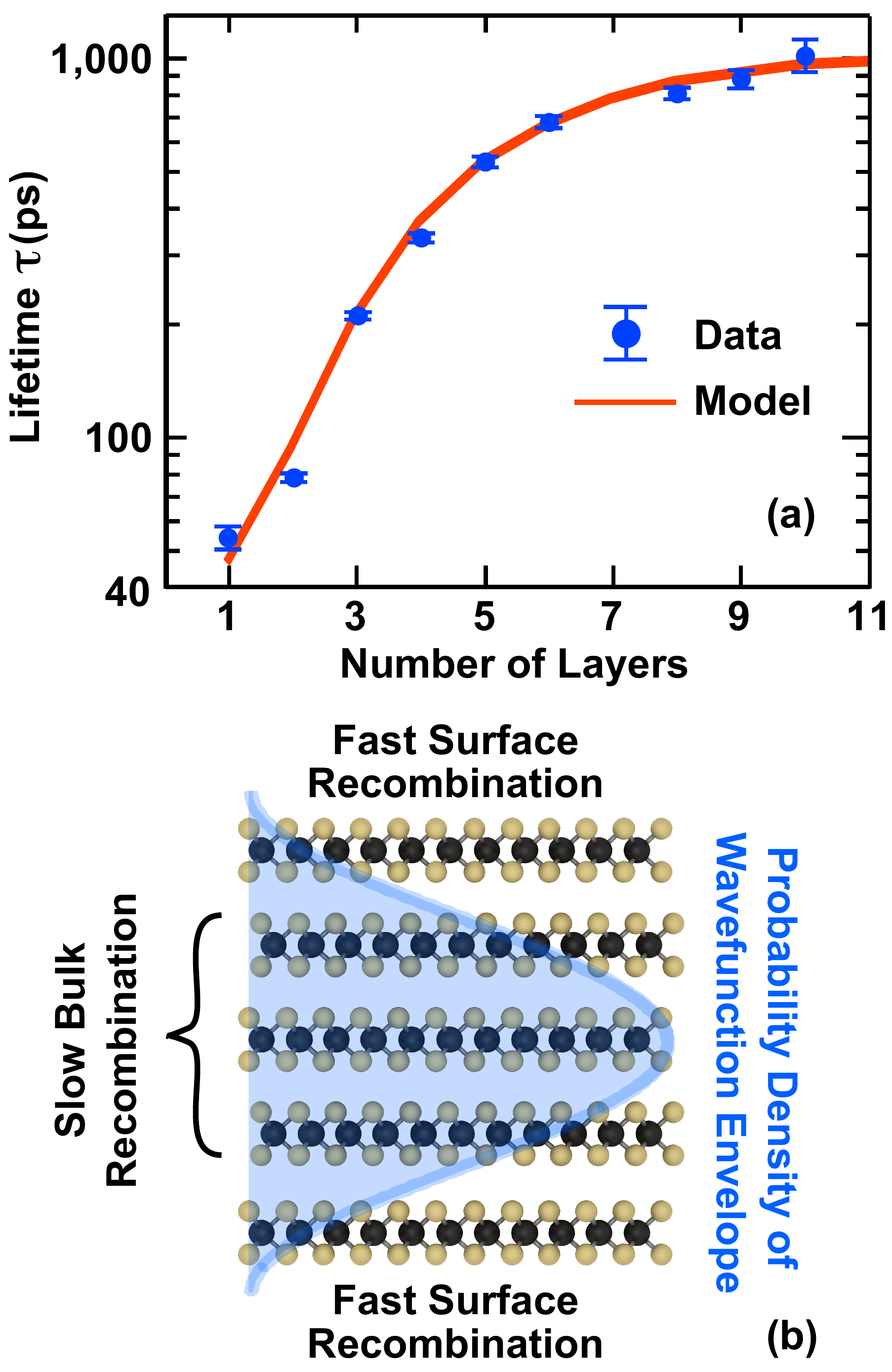}
  \caption
      {(a) (Dots) Carrier lifetime $\tau$, extracted from the data in Figure \ref{fig:OPOP2}(c), is plotted as a function of the number of layers. For each data point the pump fluence used was 32 $\mu$J/cm$^{2}$ and T=300K. (Solid lines) Theoretical model. The evolution of the carrier lifetime with the number of layers can be explained well by the competition between surface and bulk recombination, as explained in the text. The errorbars represent upper and lower bounds of the extracted lifetimes. (b) A depiction of the different recombination times in the bulk and at the surfaces of a few-layer MoS$_{2}$ sample. The solid line shows the probability density associated with the carrier envelope wavefunction in the few-layer sample.}
  \label{fig:model}
\end{figure}

\hfill \\
\noindent
\textbf{Scaling of the Carrier Lifetime with the Number of Layers:}
\newline
We now explore the dependence of the carrier lifetime $\tau$ on the number of layers. The carrier lifetime, extracted from the data shown in Figure \ref{fig:OPOP2}(c), is plotted in Figure \ref{fig:model}(a) as a function of the number of layers. The lifetime increases rapidly from $\sim$50 ps in a monolayer sample to $\sim$500 ps in a 5-layer sample and then the rate of increase slows down and the lifetime value equals $\sim$1.0 ns in a 10-layer sample. These lifetime values are consistent with the results obtained earlier by the authors for monolayer and bulk samples~\cite{Wang15,Strait14}. The question that arises is why the lifetimes vary so drastically between monolayer and multilayer samples and if any simple model can capture the seemingly complicated evolution of the lifetimes with the number of layers. Monolayer and multilayer MoS$_{2}$ samples have different bandstructures and, therefore, different conduction and valence band Bloch states for electrons and holes~\cite{Mak10,Lambrecht12,Zhao13}. However, the bandstructures of few-layer MoS$_{2}$ for different number of layers are not that different from each other when the number of layers is greater than or equal to two, and even the bandgaps of few-layer MoS$_{2}$ quickly converge to the bulk values in just 4-5 layers~\cite{Mak10}. Therefore, we rule out the variation in the measured carrier lifetimes as coming from bandstructure changes with the layer number.

A possible difference between monolayer and few-layer MoS$_2$ is the possibility of substantially more defect states, acting as recombination centers, present at the surfaces compared to in the bulk. Drastic reduction of carrier lifetimes and photoluminescence quantum efficiencies in more traditional semiconductor nanostructures, such as quantum wells and wires, due to very large surface recombination velocities is well known~\cite{Tsang88,Germann89,Moison90,Shalish04,Klimov99}. One might assume, as the simplest model, that all recombination occurs at the two surface layers in few-layer MoS$_2$ and no significant recombination occurs in the bulk. This would mean that the carriers photoexcited in the bulk would have to diffuse to the surfaces in order to recombine and carrier lifetimes would then be diffusion limited and sample thickness dependent. Diffusion time in a clean disorder-free material sample goes as the square of the sample thickness~\cite{Slutsky04} whereas the lifetimes in Figure \ref{fig:model}(a) have a very different, and much stronger, dependence on the film thickness (or the number of layers). Diffusion time in disordered materials (with correlated potential disorder) goes exponentially with the sample thickness but is also extremely temperature dependent~\cite{Slutsky04}, which is again inconsistent with our data. We propose another model here based on the quantum mechanical wavefunction of the electrons and holes in few-layer samples. Since the few-layer samples used in our work are thin (all less than 6 nm in thickness) one would expect the quantum mechanical spatial coherence of the electron wavefunction to hold in the direction perpendicular to the layers even at room temperature. We assume a fast recombination time $\tau_{s}$ (as given by the products of the Auger rate constants and the defect density) for the two surface layers in a few-layer sample, and a slow recombination time $\tau_{i}$ for all the inner layers, and then estimate the actual recombination time for a few-layer sample by weighing the inverse lifetime with the probability of electron (or hole) occupation of each layer as given by the electron (or hole) wavefunction in a few-layer sample. This procedure is equivalent to assuming a spatially varying defect structure in a more formal calculation technique such as the one presented by Wang et~al.~\cite{Wang15b}. This procedure also assumes that the carriers are mobile in the direction normal to the plane of the layers, an assumption that seems to be justified by the large splittings of the energy subbands observed in the conduction band minima and the valence band maxima in few layer MoS$_{2}$ in calculations~\cite{Ellis11}. The envelope of the electron (or hole) wavefunction of the lowest quantum confined state (confined in the direction perpendicular to the layers) can be obtained using the effective mass approximation (and assuming a discrete space),
\begin{equation}
  \psi(x,y,k) = \phi(k)\,f(x,y) = \sqrt{\frac{2}{1+N}} \sin\left(\frac{\pi k}{1+N}\right) \,f(x,y) \label{eq:scaling}
\end{equation}
where $f(x,y)$ is the envelope wavefunction in the plane of a layer, $\phi(k)$ is the envelope wavefunction in the direction perpendicular to the layers evaluated at the $k$-th layer, and $N$ is the total number of layers. The lifetime $\tau$ of the carriers in a few-layer ($N \ge 2$) sample can be estimated as,
\begin{equation}
\frac{1}{\tau} = \frac{0.5 \, (|\phi(1)|^2 + |\phi(N)|^2)}{\tau_{s}} + \frac{1 - 0.5 \, (|\phi(1)|^2 + |\phi(N)|^2)}{\tau_{i}}  
\end{equation}
The results thus obtained are plotted in Figure \ref{fig:model}(a) along with the data. We used values of $\tau_{s}$ and $\tau_{i}$ equal to 50 ps and 1.3 ns, respectively, to obtain a good fit. The excellent agreement between the model and the data for all layer numbers $N$ shows that the model captures the essential physics. Note that we only considered the lowest quantum confined state in Eq.(\ref{eq:scaling}). There is a possibility that higher energy quantum confined states in our few-layer samples could be occupied by hot or thermally excited carriers. However, even in the 10-layer sample the second confined state is estimated to be 50-100 meV higher in energy~\cite{Ellis11} and is, therefore, not expected to have significant carrier population. Also, we see no temperature dependence in the measured lifetimes and this indicates that there is no significant error made in ignoring carrier spilling into the higher quantum confined states. 

\hfill \\
\noindent
\textbf{Conclusion:}
\newline
Our results for the photoexcited carrier lifetimes in few-layer MoS$_{2}$ suggest that carrier recombination is dominated by defect-assisted processes that have much higher rates at the surface layers than in the inner layers. The excellent agreement between the data and the model for the scaling of the carrier lifetimes with the number of layers points to the validity of treating the electron and hole states in few-layer TMDs as quantum confined states describable by envelope wavefunctions in the effective mass approximation in the same way as is done in the case of more traditional semiconductor nanostructures, such as quantum wells. The exact nature of the defects that contribute the most to carrier recombination in TMDs remains unclear. Sulfur vacancies in MoS$_{2}$ were considered a strong candidate in a recent theoretical work by Wang et~al.~\cite{Wang15b} and, given the low formation energies of sulfur vacancies~\cite{Kim14}, surface layers are expected to have more of them than the inner layers in few-layer samples. Our work also shows that monolayer light emitting and detecting TMD devices will pay a high penalty in terms of the quantum efficiency unless suitable schemes for controlling and/or reducing surface defects are developed.
 
\hfill \\
\noindent
\textbf{Acknowledgments:}
\newline
The authors would like to acknowledge helpful discussions with Jared H. Strait, Michael G. Spencer, and Paul L. McEuen, and support from CCMR under NSF grant number DMR-1120296, AFOSR-MURI under grant number FA9550-09-1-0705, ONR under grant number N00014-12-1-0072, and the Cornell Center for Nanoscale Systems funded by NSF.

\bibliography{m_opop_paper}

\providecommand{\latin}[1]{#1}
\providecommand*\mcitethebibliography{\thebibliography}
\csname @ifundefined\endcsname{endmcitethebibliography}
  {\let\endmcitethebibliography\endthebibliography}{}
\begin{mcitethebibliography}{53}
\providecommand*\natexlab[1]{#1}
\providecommand*\mciteSetBstSublistMode[1]{}
\providecommand*\mciteSetBstMaxWidthForm[2]{}
\providecommand*\mciteBstWouldAddEndPuncttrue
  {\def\EndOfBibitem{\unskip.}}
\providecommand*\mciteBstWouldAddEndPunctfalse
  {\let\EndOfBibitem\relax}
\providecommand*\mciteSetBstMidEndSepPunct[3]{}
\providecommand*\mciteSetBstSublistLabelBeginEnd[3]{}
\providecommand*\EndOfBibitem{}
\mciteSetBstSublistMode{f}
\mciteSetBstMaxWidthForm{subitem}{(\alph{mcitesubitemcount})}
\mciteSetBstSublistLabelBeginEnd
  {\mcitemaxwidthsubitemform\space}
  {\relax}
  {\relax}

\bibitem[Mak \latin{et~al.}(2010)Mak, Lee, Hone, Shan, and Heinz]{Mak10}
Mak,~K.~F.; Lee,~C.; Hone,~J.; Shan,~J.; Heinz,~T.~F. \emph{Phys. Rev. Lett.}
  \textbf{2010}, \emph{105}, 136805\relax
\mciteBstWouldAddEndPuncttrue
\mciteSetBstMidEndSepPunct{\mcitedefaultmidpunct}
{\mcitedefaultendpunct}{\mcitedefaultseppunct}\relax
\EndOfBibitem
\bibitem[Splendiani \latin{et~al.}(2010)Splendiani, Sun, Zhang, Li, Kim, Chim,
  Galli, and Wang]{Splendiani10}
Splendiani,~A.; Sun,~L.; Zhang,~Y.; Li,~T.; Kim,~J.; Chim,~C.-Y.; Galli,~G.;
  Wang,~F. \emph{Nano Letters} \textbf{2010}, \emph{10}, 1271--1275, PMID:
  20229981\relax
\mciteBstWouldAddEndPuncttrue
\mciteSetBstMidEndSepPunct{\mcitedefaultmidpunct}
{\mcitedefaultendpunct}{\mcitedefaultseppunct}\relax
\EndOfBibitem
\bibitem[Wang \latin{et~al.}(2012)Wang, Kalantar-Zadeh, Kis, Coleman, and
  Strano]{Wang12}
Wang,~Q.~H.; Kalantar-Zadeh,~K.; Kis,~A.; Coleman,~J.~N.; Strano,~M.~S.
  \emph{Nature Nanotechnology} \textbf{2012}, 699--712\relax
\mciteBstWouldAddEndPuncttrue
\mciteSetBstMidEndSepPunct{\mcitedefaultmidpunct}
{\mcitedefaultendpunct}{\mcitedefaultseppunct}\relax
\EndOfBibitem
\bibitem[Mak \latin{et~al.}(2013)Mak, He, Lee, Lee, Hone, Heinz, and
  Shan]{Mak13}
Mak,~K.~F.; He,~K.; Lee,~C.; Lee,~G.~H.; Hone,~J.; Heinz,~T.~F.; Shan,~J.
  \emph{Nature Materials} \textbf{2013}, \emph{7}, 207--211\relax
\mciteBstWouldAddEndPuncttrue
\mciteSetBstMidEndSepPunct{\mcitedefaultmidpunct}
{\mcitedefaultendpunct}{\mcitedefaultseppunct}\relax
\EndOfBibitem
\bibitem[Lopez-Sanchez \latin{et~al.}(2013)Lopez-Sanchez, Lembke, Kayci,
  Radenovic, and Kis]{Lopez13}
Lopez-Sanchez,~O.; Lembke,~D.; Kayci,~M.; Radenovic,~A.; Kis,~A. \emph{Nature
  Nanotechnology} \textbf{2013}, \emph{8}, 497--501\relax
\mciteBstWouldAddEndPuncttrue
\mciteSetBstMidEndSepPunct{\mcitedefaultmidpunct}
{\mcitedefaultendpunct}{\mcitedefaultseppunct}\relax
\EndOfBibitem
\bibitem[Ross \latin{et~al.}(2014)Ross, Klement, Jones, Ghimire, Yan, Mandrus,
  Taniguchi, Watanabe, Kitamura, Yao, Cobden, and Xu]{Ross14}
Ross,~J.~S.; Klement,~P.; Jones,~A.~M.; Ghimire,~N.~J.; Yan,~J.;
  Mandrus,~D.~G.; Taniguchi,~T.; Watanabe,~K.; Kitamura,~K.; Yao,~W.;
  Cobden,~D.~H.; Xu,~X. \emph{Nature Nanotechnology} \textbf{2014}, \emph{9},
  268--272\relax
\mciteBstWouldAddEndPuncttrue
\mciteSetBstMidEndSepPunct{\mcitedefaultmidpunct}
{\mcitedefaultendpunct}{\mcitedefaultseppunct}\relax
\EndOfBibitem
\bibitem[Yin \latin{et~al.}(2012)Yin, Li, Li, Jiang, Shi, Sun, Lu, Zhang, Chen,
  and Zhang]{Hua12}
Yin,~Z.; Li,~H.; Li,~H.; Jiang,~L.; Shi,~Y.; Sun,~Y.; Lu,~G.; Zhang,~Q.;
  Chen,~X.; Zhang,~H. \emph{ACS Nano} \textbf{2012}, \emph{6}, 74--80\relax
\mciteBstWouldAddEndPuncttrue
\mciteSetBstMidEndSepPunct{\mcitedefaultmidpunct}
{\mcitedefaultendpunct}{\mcitedefaultseppunct}\relax
\EndOfBibitem
\bibitem[Sundaram \latin{et~al.}(2013)Sundaram, Engel, Lombardo, Krupke,
  Ferrari, Avouris, and Steiner]{Steiner13}
Sundaram,~R.~S.; Engel,~M.; Lombardo,~A.; Krupke,~R.; Ferrari,~A.~C.;
  Avouris,~P.; Steiner,~M. \emph{Nano Letters} \textbf{2013}, \emph{13},
  1416--1421\relax
\mciteBstWouldAddEndPuncttrue
\mciteSetBstMidEndSepPunct{\mcitedefaultmidpunct}
{\mcitedefaultendpunct}{\mcitedefaultseppunct}\relax
\EndOfBibitem
\bibitem[Baugher \latin{et~al.}(2014)Baugher, Churchill, Yang, and
  Jarillo-Herrero]{Baugher14}
Baugher,~B. W.~H.; Churchill,~H. O.~H.; Yang,~Y.; Jarillo-Herrero,~P.
  \emph{Nature Nanotechnology} \textbf{2014}, \emph{9}, 262--267\relax
\mciteBstWouldAddEndPuncttrue
\mciteSetBstMidEndSepPunct{\mcitedefaultmidpunct}
{\mcitedefaultendpunct}{\mcitedefaultseppunct}\relax
\EndOfBibitem
\bibitem[Zhang \latin{et~al.}(2014)Zhang, Wang, Chan, Manolatou, and
  Rana]{Changjian14}
Zhang,~C.; Wang,~H.; Chan,~W.; Manolatou,~C.; Rana,~F. \emph{Phys. Rev. B}
  \textbf{2014}, \emph{89}, 205436\relax
\mciteBstWouldAddEndPuncttrue
\mciteSetBstMidEndSepPunct{\mcitedefaultmidpunct}
{\mcitedefaultendpunct}{\mcitedefaultseppunct}\relax
\EndOfBibitem
\bibitem[Wu \latin{et~al.}(2015)Wu, Buckley, Schaibley, Feng, Yan, Mandrus,
  Hatami, Yao, Vučković, Majumdar, and Xu]{Sanfeng15}
Wu,~S.; Buckley,~S.; Schaibley,~J.~R.; Feng,~L.; Yan,~J.; Mandrus,~D.~G.;
  Hatami,~F.; Yao,~W.; Vučković,~J.; Majumdar,~A.; Xu,~X. \emph{Nature}
  \textbf{2015}, \emph{520}, 69--72\relax
\mciteBstWouldAddEndPuncttrue
\mciteSetBstMidEndSepPunct{\mcitedefaultmidpunct}
{\mcitedefaultendpunct}{\mcitedefaultseppunct}\relax
\EndOfBibitem
\bibitem[Strait \latin{et~al.}(2014)Strait, Nene, and Rana]{Strait14}
Strait,~J.~H.; Nene,~P.; Rana,~F. \emph{Phys. Rev. B} \textbf{2014}, \emph{90},
  245402\relax
\mciteBstWouldAddEndPuncttrue
\mciteSetBstMidEndSepPunct{\mcitedefaultmidpunct}
{\mcitedefaultendpunct}{\mcitedefaultseppunct}\relax
\EndOfBibitem
\bibitem[Shi \latin{et~al.}(2013)Shi, Yan, Bertolazzi, Brivio, Gao, Kis, Jena,
  Xing, and Huang]{Huang13}
Shi,~H.; Yan,~R.; Bertolazzi,~S.; Brivio,~J.; Gao,~B.; Kis,~A.; Jena,~D.;
  Xing,~H.~G.; Huang,~L. \emph{ACS Nano} \textbf{2013}, \emph{7},
  1072--1080\relax
\mciteBstWouldAddEndPuncttrue
\mciteSetBstMidEndSepPunct{\mcitedefaultmidpunct}
{\mcitedefaultendpunct}{\mcitedefaultseppunct}\relax
\EndOfBibitem
\bibitem[Yu \latin{et~al.}(2013)Yu, Liu, Zhou, Yin, Li, Yu, and Duan]{Yu13}
Yu,~W.-J.; Liu,~Y.; Zhou,~H.; Yin,~A.; Li,~Z.; Yu,~H.; Duan,~X. \emph{Nature
  Nanotechnology} \textbf{2013}, \emph{8}, 952\relax
\mciteBstWouldAddEndPuncttrue
\mciteSetBstMidEndSepPunct{\mcitedefaultmidpunct}
{\mcitedefaultendpunct}{\mcitedefaultseppunct}\relax
\EndOfBibitem
\bibitem[Ma \latin{et~al.}(2014)Ma, Hu, Zhao, Tang, Wu, Zhou, and Zhang]{Ma14}
Ma,~Z.; Hu,~Z.; Zhao,~X.; Tang,~Q.; Wu,~D.; Zhou,~Z.; Zhang,~L. \emph{The
  Journal of Physical Chemistry C} \textbf{2014}, \emph{118}, 5593--5599\relax
\mciteBstWouldAddEndPuncttrue
\mciteSetBstMidEndSepPunct{\mcitedefaultmidpunct}
{\mcitedefaultendpunct}{\mcitedefaultseppunct}\relax
\EndOfBibitem
\bibitem[Ceballos \latin{et~al.}(2014)Ceballos, Bellus, Chiu, and Zhao]{Hui14}
Ceballos,~F.; Bellus,~M.~Z.; Chiu,~H.-Y.; Zhao,~H. \emph{ACS Nano}
  \textbf{2014}, \emph{8}, 12717--12724, PMID: 25402669\relax
\mciteBstWouldAddEndPuncttrue
\mciteSetBstMidEndSepPunct{\mcitedefaultmidpunct}
{\mcitedefaultendpunct}{\mcitedefaultseppunct}\relax
\EndOfBibitem
\bibitem[Cho \latin{et~al.}(2014)Cho, Kim, Park, Park, Kim, Jang, Jeong, Hong,
  and Lee]{Cho14}
Cho,~K.; Kim,~T.-Y.; Park,~W.; Park,~J.; Kim,~D.; Jang,~J.; Jeong,~H.;
  Hong,~S.; Lee,~T. \emph{Nanotechnology} \textbf{2014}, \emph{25},
  155201\relax
\mciteBstWouldAddEndPuncttrue
\mciteSetBstMidEndSepPunct{\mcitedefaultmidpunct}
{\mcitedefaultendpunct}{\mcitedefaultseppunct}\relax
\EndOfBibitem
\bibitem[Wang \latin{et~al.}(2015)Wang, Zhang, and Rana]{Wang15}
Wang,~H.; Zhang,~C.; Rana,~F. \emph{Nano Letters} \textbf{2015}, \emph{15},
  339--345, PMID: 25546602\relax
\mciteBstWouldAddEndPuncttrue
\mciteSetBstMidEndSepPunct{\mcitedefaultmidpunct}
{\mcitedefaultendpunct}{\mcitedefaultseppunct}\relax
\EndOfBibitem
\bibitem[Wang \latin{et~al.}(2012)Wang, Ruzicka, Kumar, Bellus, Chiu, and
  Zhao]{WangR12}
Wang,~R.; Ruzicka,~B.~A.; Kumar,~N.; Bellus,~M.~Z.; Chiu,~H.-Y.; Zhao,~H.
  \emph{Phys. Rev. B} \textbf{2012}, \emph{86}, 045406\relax
\mciteBstWouldAddEndPuncttrue
\mciteSetBstMidEndSepPunct{\mcitedefaultmidpunct}
{\mcitedefaultendpunct}{\mcitedefaultseppunct}\relax
\EndOfBibitem
\bibitem[Sim \latin{et~al.}(2013)Sim, Park, Song, In, Lee, Kim, and
  Choi]{Choi13}
Sim,~S.; Park,~J.; Song,~J.-G.; In,~C.; Lee,~Y.-S.; Kim,~H.; Choi,~H.
  \emph{Phys. Rev. B} \textbf{2013}, \emph{88}, 075434\relax
\mciteBstWouldAddEndPuncttrue
\mciteSetBstMidEndSepPunct{\mcitedefaultmidpunct}
{\mcitedefaultendpunct}{\mcitedefaultseppunct}\relax
\EndOfBibitem
\bibitem[Sun \latin{et~al.}(2014)Sun, Rao, Reider, Chen, You, Brézin,
  Harutyunyan, and Heinz]{Sun14}
Sun,~D.; Rao,~Y.; Reider,~G.~A.; Chen,~G.; You,~Y.; Brézin,~L.;
  Harutyunyan,~A.~R.; Heinz,~T.~F. \emph{Nano Letters} \textbf{2014},
  \emph{14}, 5625--5629\relax
\mciteBstWouldAddEndPuncttrue
\mciteSetBstMidEndSepPunct{\mcitedefaultmidpunct}
{\mcitedefaultendpunct}{\mcitedefaultseppunct}\relax
\EndOfBibitem
\bibitem[Lagarde \latin{et~al.}(2014)Lagarde, Bouet, Marie, Zhu, Liu, Amand,
  Tan, and Urbaszek]{Lagarde14}
Lagarde,~D.; Bouet,~L.; Marie,~X.; Zhu,~C.~R.; Liu,~B.~L.; Amand,~T.;
  Tan,~P.~H.; Urbaszek,~B. \emph{Phys. Rev. Lett.} \textbf{2014}, \emph{112},
  047401\relax
\mciteBstWouldAddEndPuncttrue
\mciteSetBstMidEndSepPunct{\mcitedefaultmidpunct}
{\mcitedefaultendpunct}{\mcitedefaultseppunct}\relax
\EndOfBibitem
\bibitem[Korn \latin{et~al.}(2011)Korn, Heydrich, Hirmer, Schmutzler, and
  Schüller]{Schuller11}
Korn,~T.; Heydrich,~S.; Hirmer,~M.; Schmutzler,~J.; Schüller,~C. \emph{Applied
  Physics Letters} \textbf{2011}, \emph{99}, 102109\relax
\mciteBstWouldAddEndPuncttrue
\mciteSetBstMidEndSepPunct{\mcitedefaultmidpunct}
{\mcitedefaultendpunct}{\mcitedefaultseppunct}\relax
\EndOfBibitem
\bibitem[Docherty \latin{et~al.}(2014)Docherty, Parkinson, Joyce, Chiu, Chen,
  Lee, Li, Herz, and Johnston]{Docherty14}
Docherty,~C.~J.; Parkinson,~P.; Joyce,~H.~J.; Chiu,~M.-H.; Chen,~C.-H.;
  Lee,~M.-Y.; Li,~L.-J.; Herz,~L.~M.; Johnston,~M.~B. \emph{ACS Nano}
  \textbf{2014}, \emph{8}, 11147--11153, PMID: 25347405\relax
\mciteBstWouldAddEndPuncttrue
\mciteSetBstMidEndSepPunct{\mcitedefaultmidpunct}
{\mcitedefaultendpunct}{\mcitedefaultseppunct}\relax
\EndOfBibitem
\bibitem[Wang \latin{et~al.}(2015)Wang, Strait, Zhang, Chan, Manolatou, Tiwari,
  and Rana]{Wang15b}
Wang,~H.; Strait,~J.~H.; Zhang,~C.; Chan,~W.; Manolatou,~C.; Tiwari,~S.;
  Rana,~F. \emph{Phys. Rev. B} \textbf{2015}, \emph{91}, 165411\relax
\mciteBstWouldAddEndPuncttrue
\mciteSetBstMidEndSepPunct{\mcitedefaultmidpunct}
{\mcitedefaultendpunct}{\mcitedefaultseppunct}\relax
\EndOfBibitem
\bibitem[Fuhr \latin{et~al.}(2004)Fuhr, Sa\'ul, and Sofo]{Sofo04}
Fuhr,~J.~D.; Sa\'ul,~A.; Sofo,~J.~O. \emph{Phys. Rev. Lett.} \textbf{2004},
  \emph{92}, 026802\relax
\mciteBstWouldAddEndPuncttrue
\mciteSetBstMidEndSepPunct{\mcitedefaultmidpunct}
{\mcitedefaultendpunct}{\mcitedefaultseppunct}\relax
\EndOfBibitem
\bibitem[Komsa \latin{et~al.}(2012)Komsa, Kotakoski, Kurasch, Lehtinen, Kaiser,
  and Krasheninnikov]{Komsa12}
Komsa,~H.-P.; Kotakoski,~J.; Kurasch,~S.; Lehtinen,~O.; Kaiser,~U.;
  Krasheninnikov,~A.~V. \emph{Phys. Rev. Lett.} \textbf{2012}, \emph{109},
  035503\relax
\mciteBstWouldAddEndPuncttrue
\mciteSetBstMidEndSepPunct{\mcitedefaultmidpunct}
{\mcitedefaultendpunct}{\mcitedefaultseppunct}\relax
\EndOfBibitem
\bibitem[Enyashin \latin{et~al.}(2013)Enyashin, Bar-Sadan, Houben, and
  Seifert]{Seifert13}
Enyashin,~A.~N.; Bar-Sadan,~M.; Houben,~L.; Seifert,~G. \emph{The Journal of
  Physical Chemistry C} \textbf{2013}, \emph{117}, 10842--10848\relax
\mciteBstWouldAddEndPuncttrue
\mciteSetBstMidEndSepPunct{\mcitedefaultmidpunct}
{\mcitedefaultendpunct}{\mcitedefaultseppunct}\relax
\EndOfBibitem
\bibitem[Zhou \latin{et~al.}(2013)Zhou, Zou, Najmaei, Liu, Shi, Kong, Lou,
  Ajayan, Yakobson, and Idrobo]{Zhou13}
Zhou,~W.; Zou,~X.; Najmaei,~S.; Liu,~Z.; Shi,~Y.; Kong,~J.; Lou,~J.;
  Ajayan,~P.~M.; Yakobson,~B.~I.; Idrobo,~J.-C. \emph{Nano Letters}
  \textbf{2013}, \emph{13}, 2615--2622\relax
\mciteBstWouldAddEndPuncttrue
\mciteSetBstMidEndSepPunct{\mcitedefaultmidpunct}
{\mcitedefaultendpunct}{\mcitedefaultseppunct}\relax
\EndOfBibitem
\bibitem[Noh \latin{et~al.}(2014)Noh, Kim, and Kim]{Kim14}
Noh,~J.-Y.; Kim,~H.; Kim,~Y.-S. \emph{Phys. Rev. B} \textbf{2014}, \emph{89},
  205417\relax
\mciteBstWouldAddEndPuncttrue
\mciteSetBstMidEndSepPunct{\mcitedefaultmidpunct}
{\mcitedefaultendpunct}{\mcitedefaultseppunct}\relax
\EndOfBibitem
\bibitem[Yuan \latin{et~al.}(2014)Yuan, Rold\'an, Katsnelson, and
  Guinea]{Guinea14}
Yuan,~S.; Rold\'an,~R.; Katsnelson,~M.~I.; Guinea,~F. \emph{Phys. Rev. B}
  \textbf{2014}, \emph{90}, 041402\relax
\mciteBstWouldAddEndPuncttrue
\mciteSetBstMidEndSepPunct{\mcitedefaultmidpunct}
{\mcitedefaultendpunct}{\mcitedefaultseppunct}\relax
\EndOfBibitem
\bibitem[Liu \latin{et~al.}(2013)Liu, Guo, Fang, and Robertson]{Robertson13}
Liu,~D.; Guo,~Y.; Fang,~L.; Robertson,~J. \emph{Applied Physics Letters}
  \textbf{2013}, \emph{103}, 183113\relax
\mciteBstWouldAddEndPuncttrue
\mciteSetBstMidEndSepPunct{\mcitedefaultmidpunct}
{\mcitedefaultendpunct}{\mcitedefaultseppunct}\relax
\EndOfBibitem
\bibitem[Qiu \latin{et~al.}(2013)Qiu, Xu, Wang, Ren, Nan, Ni, Chen, Yuan, Miao,
  Song, Long, Shi, Sun, Wang, and Wang]{Hao13}
Qiu,~H.; Xu,~T.; Wang,~Z.; Ren,~W.; Nan,~H.; Ni,~Z.; Chen,~Q.; Yuan,~S.;
  Miao,~F.; Song,~F.; Long,~G.; Shi,~Y.; Sun,~L.; Wang,~J.; Wang,~X.
  \emph{Nature Communications} \textbf{2013}, \emph{4}, --\relax
\mciteBstWouldAddEndPuncttrue
\mciteSetBstMidEndSepPunct{\mcitedefaultmidpunct}
{\mcitedefaultendpunct}{\mcitedefaultseppunct}\relax
\EndOfBibitem
\bibitem[Van~der Zande \latin{et~al.}(2013)Van~der Zande, Huang, Chenet,
  Berkelbach, You, Lee, Heinz, Reichman, Muller, and Hone]{VanDerZande13}
Van~der Zande,~A.~M.; Huang,~P.~Y.; Chenet,~D.~A.; Berkelbach,~T.~C.; You,~Y.;
  Lee,~G.-H.; Heinz,~T.~F.; Reichman,~D.~R.; Muller,~D.~A.; Hone,~J.~C.
  \emph{Nature Materials} \textbf{2013}, 554--561\relax
\mciteBstWouldAddEndPuncttrue
\mciteSetBstMidEndSepPunct{\mcitedefaultmidpunct}
{\mcitedefaultendpunct}{\mcitedefaultseppunct}\relax
\EndOfBibitem
\bibitem[Furchi \latin{et~al.}(2014)Furchi, Polyushkin, Pospischil, and
  Mueller]{Marco14}
Furchi,~M.~M.; Polyushkin,~D.~K.; Pospischil,~A.; Mueller,~T. \emph{Nano
  Letters} \textbf{2014}, \emph{14}, 6165--6170\relax
\mciteBstWouldAddEndPuncttrue
\mciteSetBstMidEndSepPunct{\mcitedefaultmidpunct}
{\mcitedefaultendpunct}{\mcitedefaultseppunct}\relax
\EndOfBibitem
\bibitem[Tai \latin{et~al.}(1988)Tai, Hayes, McCall, and Tsang]{Tsang88}
Tai,~K.; Hayes,~T.~R.; McCall,~S.~L.; Tsang,~W.~T. \emph{Applied Physics
  Letters} \textbf{1988}, \emph{53}, 302--303\relax
\mciteBstWouldAddEndPuncttrue
\mciteSetBstMidEndSepPunct{\mcitedefaultmidpunct}
{\mcitedefaultendpunct}{\mcitedefaultseppunct}\relax
\EndOfBibitem
\bibitem[Maile \latin{et~al.}(1989)Maile, Forchel, Germann, and
  Grützmacher]{Germann89}
Maile,~B.~E.; Forchel,~A.; Germann,~R.; Grützmacher,~D. \emph{Applied Physics
  Letters} \textbf{1989}, \emph{54}, 1552--1554\relax
\mciteBstWouldAddEndPuncttrue
\mciteSetBstMidEndSepPunct{\mcitedefaultmidpunct}
{\mcitedefaultendpunct}{\mcitedefaultseppunct}\relax
\EndOfBibitem
\bibitem[Moison \latin{et~al.}(1990)Moison, Elcess, Houzay, Marzin, G\'erard,
  Barthe, and Bensoussan]{Moison90}
Moison,~J.~M.; Elcess,~K.; Houzay,~F.; Marzin,~J.~Y.; G\'erard,~J.~M.;
  Barthe,~F.; Bensoussan,~M. \emph{Phys. Rev. B} \textbf{1990}, \emph{41},
  12945--12948\relax
\mciteBstWouldAddEndPuncttrue
\mciteSetBstMidEndSepPunct{\mcitedefaultmidpunct}
{\mcitedefaultendpunct}{\mcitedefaultseppunct}\relax
\EndOfBibitem
\bibitem[Shalish \latin{et~al.}(2004)Shalish, Temkin, and
  Narayanamurti]{Shalish04}
Shalish,~I.; Temkin,~H.; Narayanamurti,~V. \emph{Phys. Rev. B} \textbf{2004},
  \emph{69}, 245401\relax
\mciteBstWouldAddEndPuncttrue
\mciteSetBstMidEndSepPunct{\mcitedefaultmidpunct}
{\mcitedefaultendpunct}{\mcitedefaultseppunct}\relax
\EndOfBibitem
\bibitem[Klimov \latin{et~al.}(1999)Klimov, McBranch, Leatherdale, and
  Bawendi]{Klimov99}
Klimov,~V.~I.; McBranch,~D.~W.; Leatherdale,~C.~A.; Bawendi,~M.~G. \emph{Phys.
  Rev. B} \textbf{1999}, \emph{60}, 13740--13749\relax
\mciteBstWouldAddEndPuncttrue
\mciteSetBstMidEndSepPunct{\mcitedefaultmidpunct}
{\mcitedefaultendpunct}{\mcitedefaultseppunct}\relax
\EndOfBibitem
\bibitem[Yablonovitch \latin{et~al.}(1986)Yablonovitch, Allara, Chang, Gmitter,
  and Bright]{Bright86}
Yablonovitch,~E.; Allara,~D.~L.; Chang,~C.~C.; Gmitter,~T.; Bright,~T.~B.
  \emph{Phys. Rev. Lett.} \textbf{1986}, \emph{57}, 249--252\relax
\mciteBstWouldAddEndPuncttrue
\mciteSetBstMidEndSepPunct{\mcitedefaultmidpunct}
{\mcitedefaultendpunct}{\mcitedefaultseppunct}\relax
\EndOfBibitem
\bibitem[Wang \latin{et~al.}(2013)Wang, Lin, Chen, Chang, Huang, Yang,
  Tjahjono, Huang, Hsu, and Chen]{Wang13}
Wang,~W.-C.; Lin,~C.-W.; Chen,~H.-J.; Chang,~C.-W.; Huang,~J.-J.; Yang,~M.-J.;
  Tjahjono,~B.; Huang,~J.-J.; Hsu,~W.-C.; Chen,~M.-J. \emph{ACS Applied
  Materials \& Interfaces} \textbf{2013}, \emph{5}, 9752--9759, PMID:
  24028609\relax
\mciteBstWouldAddEndPuncttrue
\mciteSetBstMidEndSepPunct{\mcitedefaultmidpunct}
{\mcitedefaultendpunct}{\mcitedefaultseppunct}\relax
\EndOfBibitem
\bibitem[Dan \latin{et~al.}(2011)Dan, Seo, Takei, Meza, Javey, and
  Crozier]{Dan11}
Dan,~Y.; Seo,~K.; Takei,~K.; Meza,~J.~H.; Javey,~A.; Crozier,~K.~B. \emph{Nano
  Letters} \textbf{2011}, \emph{11}, 2527--2532, PMID: 21598980\relax
\mciteBstWouldAddEndPuncttrue
\mciteSetBstMidEndSepPunct{\mcitedefaultmidpunct}
{\mcitedefaultendpunct}{\mcitedefaultseppunct}\relax
\EndOfBibitem
\bibitem[Lee \latin{et~al.}(2010)Lee, Yan, Brus, Heinz, Hone, and Ryu]{Ryu10}
Lee,~C.; Yan,~H.; Brus,~L.~E.; Heinz,~T.~F.; Hone,~J.; Ryu,~S. \emph{ACS Nano}
  \textbf{2010}, \emph{4}, 2695--2700, PMID: 20392077\relax
\mciteBstWouldAddEndPuncttrue
\mciteSetBstMidEndSepPunct{\mcitedefaultmidpunct}
{\mcitedefaultendpunct}{\mcitedefaultseppunct}\relax
\EndOfBibitem
\bibitem[Chakraborty \latin{et~al.}(2012)Chakraborty, Bera, Muthu, Bhowmick,
  Waghmare, and Sood]{Sood12}
Chakraborty,~B.; Bera,~A.; Muthu,~D. V.~S.; Bhowmick,~S.; Waghmare,~U.~V.;
  Sood,~A.~K. \emph{Phys. Rev. B} \textbf{2012}, \emph{85}, 161403\relax
\mciteBstWouldAddEndPuncttrue
\mciteSetBstMidEndSepPunct{\mcitedefaultmidpunct}
{\mcitedefaultendpunct}{\mcitedefaultseppunct}\relax
\EndOfBibitem
\bibitem[Zhao \latin{et~al.}(2013)Zhao, Ribeiro, Toh, Carvalho, Kloc,
  Castro~Neto, and Eda]{Zhao13}
Zhao,~W.; Ribeiro,~R.~M.; Toh,~M.; Carvalho,~A.; Kloc,~C.; Castro~Neto,~A.~H.;
  Eda,~G. \emph{Nano Letters} \textbf{2013}, \emph{13}, 5627--5634, PMID:
  24168432\relax
\mciteBstWouldAddEndPuncttrue
\mciteSetBstMidEndSepPunct{\mcitedefaultmidpunct}
{\mcitedefaultendpunct}{\mcitedefaultseppunct}\relax
\EndOfBibitem
\bibitem[Ellis \latin{et~al.}(2011)Ellis, Lucero, and Scuseria]{Ellis11}
Ellis,~J.~K.; Lucero,~M.~J.; Scuseria,~G.~E. \emph{Applied Physics Letters}
  \textbf{2011}, \emph{99}, 261908\relax
\mciteBstWouldAddEndPuncttrue
\mciteSetBstMidEndSepPunct{\mcitedefaultmidpunct}
{\mcitedefaultendpunct}{\mcitedefaultseppunct}\relax
\EndOfBibitem
\bibitem[Fan(1967)]{scsm67}
Fan,~H.~Y. \emph{Semiconductors and Semimetals}; Academic Press: New York, USA,
  1967; Vol.~3\relax
\mciteBstWouldAddEndPuncttrue
\mciteSetBstMidEndSepPunct{\mcitedefaultmidpunct}
{\mcitedefaultendpunct}{\mcitedefaultseppunct}\relax
\EndOfBibitem
\bibitem[Buus and Amann(1998)Buus, and Amann]{tld98}
Buus,~J.; Amann,~M. \emph{Tunebale Laser diodes}, 1st ed.; Artech House
  Publishers: New York, USA, 1998\relax
\mciteBstWouldAddEndPuncttrue
\mciteSetBstMidEndSepPunct{\mcitedefaultmidpunct}
{\mcitedefaultendpunct}{\mcitedefaultseppunct}\relax
\EndOfBibitem
\bibitem[Jellison and Modine(1994)Jellison, and Modine]{Modine94}
Jellison,~G.~E.; Modine,~F.~A. \emph{Journal of Applied Physics} \textbf{1994},
  \emph{76}, 3758--3761\relax
\mciteBstWouldAddEndPuncttrue
\mciteSetBstMidEndSepPunct{\mcitedefaultmidpunct}
{\mcitedefaultendpunct}{\mcitedefaultseppunct}\relax
\EndOfBibitem
\bibitem[Cheiwchanchamnangij and Lambrecht(2012)Cheiwchanchamnangij, and
  Lambrecht]{Lambrecht12}
Cheiwchanchamnangij,~T.; Lambrecht,~W. R.~L. \emph{Phys. Rev. B} \textbf{2012},
  \emph{85}, 205302\relax
\mciteBstWouldAddEndPuncttrue
\mciteSetBstMidEndSepPunct{\mcitedefaultmidpunct}
{\mcitedefaultendpunct}{\mcitedefaultseppunct}\relax
\EndOfBibitem
\bibitem[Slutsky \latin{et~al.}(2004)Slutsky, Kardar, and Mirny]{Slutsky04}
Slutsky,~M.; Kardar,~M.; Mirny,~L.~A. \emph{Phys. Rev. E} \textbf{2004},
  \emph{69}, 061903\relax
\mciteBstWouldAddEndPuncttrue
\mciteSetBstMidEndSepPunct{\mcitedefaultmidpunct}
{\mcitedefaultendpunct}{\mcitedefaultseppunct}\relax
\EndOfBibitem
\end{mcitethebibliography}

\newpage
\clearpage
\includepdf[pages={{},-}]{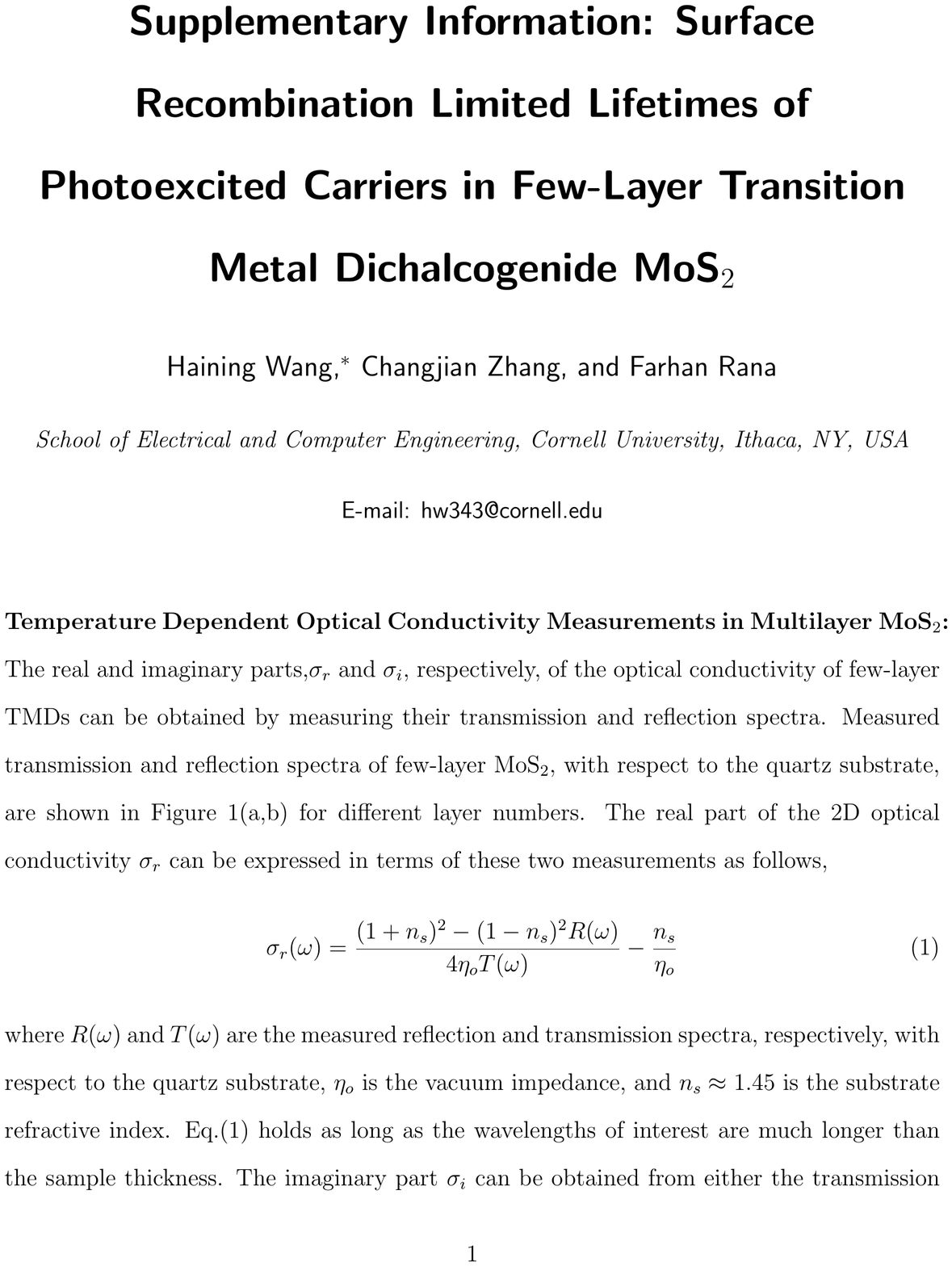}

\end{document}